\documentclass[galaxies,review,accept,moreauthors,pdftex,galaxies]{mdpi}

\firstpage{1}
\pubvolume{xx}
\issuenum{1}
\articlenumber{5}
\pubyear{2019}
\copyrightyear{2019}
\history{Received: 28 February 2019; Accepted: 8 April 2019; Published: date}
\updates{yes} 

\Title{The RoboPol Program: Optical Polarimetric Monitoring of Blazars}

\Author{Dmitry Blinov$^{1,2,3,*}$\orcidA{} and Vasiliki Pavlidou $^{1,2,*}$\orcidB{} for the RoboPol 
collaboration}

\AuthorNames{Firstname Lastname, Firstname Lastname and Firstname Lastname}

\address{%
$^{1}$ \quad Physics Department \& Institute of Theoretical \& Computational Physics, University of Crete, 71003~Heraklion,~Crete,~Greece\\
$^{2}$ \quad IESL \& Institute of Astrophysics, Foundation for Research and Technology-Hellas, 71110~Heraklion,~Crete,~Greece\\
$^{3}$ \quad Astronomical Institute, Saint Petersburg State University, 198504 Saint Petersburg, 
Russia}

\corres{Correspondence: blinov@physics.uoc.gr (D.B.); pavlidou@physics.uoc.gr (V.P.)}

\abstract{After three years of polarimetric monitoring of blazars, the~RoboPol project has
uncovered several key characteristics of polarimetric rotations in the optical for these most
variable sources. The~most important of these is that polarization properties of the synchrotron
emission in the optical appear to be directly linked with gamma-ray activity. In this paper, we discuss the
evidence for this connection, as~well as the broader features of polarimetric behavior in blazars
that are key in making progress with theoretical modeling of blazar emission.}


\newcommand{\dg}{$^{\circ}$\,}
\newcommand{\g}{$\gamma$}

\begin{document}

\section{Introduction} \label{sec:intr}
The first measurements of optical polarization of quasars followed soon after the identification of their
optical counterparts. These measurements have revealed that many of these sources are linearly
polarized to some extent, and variable in both the degree of polarization and the electric
vector position angle (EVPA) \citep{Kinman1966}. Fifteen years later, optical polarization was
measured for $\sim$60 compact extragalactic sources, and the first statistical conclusions could be
derived, together with some preliminary tests of consistency with models \citep{Angel1980}. It was found that,
typically, the behavior of polarization degrees and angles are erratic \citep{Moore1982}. This can be
explained by the presence of a persistent component and one or multiple variable components
\citep{Brindle1985,Blinov2009,Uemura2010,Villforth2010}. In fact, such a model was first used for the
interpretation of irregular radio polarization behavior as early as 1967 \citep{Goldstein1967}.
However, it has been known since the end of the 1970s that some sources exhibit long monotonic
rotations of the EVPA in the radio band emission during flares \citep{Ledden1979,Jones1985}. The
first optical polarization rotation, simultaneous with a similar rotation in the radio band, was
reported by Kikuchi et al. (1988) \citep{Kikuchi1988}. During the first years of operation of the
Fermi \g-ray observatory, optical EVPA rotations have attracted a lot of interest, because it
was noticed that some of the strong \g-ray flares were accompanied by these events
\citep{Marscher2008,Abdo2010,Marscher2010}. If EVPA rotations were physically related to the 
high-energy activity, they could help localize and clarify the nature of the latter one, which was
actively discussed at that time.

Therefore, a number of optical polarization monitoring programs have since focused on observations 
of blazars that are bright and active in \g-rays. Such major monitoring programs have been carried 
out by the Boston University group \citep{Jorstad2016}, at St. Petersburg University 
\citep{Hagen2013}, at the Hiroshima University  
\textcolor{black}{(Kanata~program)}~\citep{Itoh2018}, at the Steward Observatory \cite{Smith2009}, 
at the Calar Alto observatory \citep{Agudo2012}, and at the Tuorla observatory \citep{Takalo2008}.

From a theoretical standpoint, a~wide variety of mechanisms have been proposed to explain the
origin of EVPA rotations. They include the following scenarios:
\begin{enumerate}[leftmargin=10mm,labelsep=5.8mm]
\item A two-component model, where changes in the polarization are due only to changes in the
relative flux of the two components \citep{Bjornsson1982}.
\item Propagation of an emission component along the jet with a non-axisymmetric magnetic field
\citep{Konigl1985}.
\item \textcolor{black}{Polarized synchrotron flares rotating in the accretion disc}
\citep{Sillanpaa1993}.
\item Motion of an emission component along a helical streamline in the helical magnetic field
\citep{Marscher2008,Marscher2010}.
\item Jet bending \citep{Abdo2010}.
\item Changes of polarization parameters and total flux due to the Doppler factor variability
\citep{Larionov2013}.
\textls[-15]{\item Random variation of the polarization vector in the emission produced by a turbulent~zone~\citep{Jones1985,Marscher2014,Kiehlmann2016}.}
\item \textcolor{black}{Propagation of a relativistic shock in the jet that causes compression of the
magnetic field and alters its direction and the degree of ordering} \citep{Zhang2014}.
\item Precession of the jet \citep{Lyutikov2017}.
\item Kink instability \citep{Nalewajko2017}.
\item Magnetic reconnections \citep{Zhang2018}.
\item Relativistic EVPA rotation in the observer frame combined with the conical jet geometry and
particle cooling \citep{Peirson2018}.
\end{enumerate}

Inspired by the perplexing and unclear nature of EVPA rotations, an intensive polarization
monitoring program, RoboPol, was designed and executed.

\section{The RoboPol Program}

In spite of the immense significance of EVPA rotations for an understanding of the magnetic field
structure in the jets and their importance for constraining theoretical models, only a handful of
these events were reported in the 25 years since their discovery. \textcolor{black}{Additionally,
most of these observing programs} were monitoring small numbers of the most violently variable blazars;
these were biased samples, and not appropriate for statistical studies. Therefore, such questions as
"Do all blazars rotate the EVPA? How often these rotations  occur on average? Does this depend on
the blazar class?" would be impossible to address accurately using data of monitoring programs
already existing at the time when RoboPol~was~designed.

In order to overcome these shortcomings, an international collaboration between the University of
Crete and Foundation of Research and Technology Hellas (FORTH, Greece), the~Max-Planck Institute for
Radioastronomy (Germany), the~California Institute of Technology (USA), the~Inter-University Centre
for Astronomy and Astrophysics (India), and the Nicolaus Copernicus University (Poland) developed
a new blazar monitoring program called RoboPol. This program specifically aimed at maximizing the
number of detected EVPA rotations and characterizing the optical polarization behavior of blazars.

\subsection{Instrumentation}

\begin{figure}[H]
 \centering
 \includegraphics[width=0.45\textwidth]{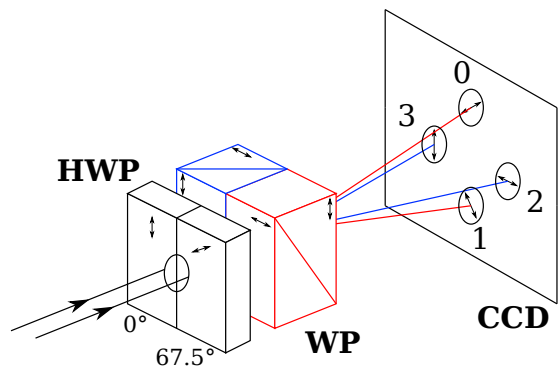}
 \caption{Schematic of the polarizing optics of RoboPol. The~array of two half-wave plates
(HWP) and two Wollaston prisms (WP) is shown. The~fast optical axes of the elements and the
polarization planes of the rays are shown with arrows. \textcolor{black}{The HWP with the fast axis
oriented at 67.5\dg rotates the polarization plane by $2\cdot67.5 = 135$\dg (see, e.g.,
\cite{clarke2009}).}}
 \label{fig:optics}
\end{figure}

The key element of the the RoboPol program is a novel-design optical polarimeter that was developed
specifically for the project. This polarimeter has a fixed set of half wave plates and Wollaston
prisms (see Figure \ref{fig:optics}) that split each incident ray into four rays with the polarization
plane rotated by 45\dg with respect to each other. Therefore, one can measure the Stokes parameters
$q = Q/I$ and $u = U/I$ of the linearly polarized light in one exposure. Given the large field of
view (13'~$\times$~13'), the total flux density $I$ can also be measured at the same time if a
comparison star with a known magnitude is present in the field. Since the polarimeter has no moving
parts except the filter, it avoids systematic uncertainties related to an imperfect alignment of
rotating optical elements, and simplifies the~instrument~calibration.

The polarimeter was hosted on a 1.3 m telescope at the Skinakas Observatory, which is a joint
facility of the FORTH and the University of Crete. It is located at an altitude of 1750~m in Crete
(35.2120\dg N, 24.8982\dg E).

The site offers seeing conditions which are among the best in Europe, comparable to several of the largest
facilities in the world (median seeing of 0.7 arcsec \citep{Boumis2001}). The~Skinakas Observatory
granted the RoboPol collaboration four nights per week, on average, over the entire Skinakas
observing season (May~to~November~each~year), for three years (2013--2015).

In addition to the advanced instrument, we developed an automated telescope control system, as well as a
processing pipeline. The~former one was able to point automatically and estimate the brightness of
targets using the pointing exposures. It could also estimate polarization parameters on the fly,
reaching a particular signal-to-noise ratio in EVPA if required.

A more detailed description of the polarimeter can be found in Ramaprakash et al. (2019)
\cite{Ram2019}, while the pipeline and the control system are presented in detail in King et
al. (2014) \citep{King2014}.

\subsection{Unbiased Observing Sample}

\textls[-5]{Another unique feature of the program is the observing sample, constructed in a way which allows for
systematic statistical studies. We built the sample relying on unbiased and statistically rigorous
criteria. Firstly, we constructed a flux-limited sample of \g--loud blazars from the second
Fermi-LAT (2FGL) source catalog \citep{Nolan2012}, selecting sources classified as BL Lac (bzb),
FSRQ (bzq), or an active galaxy of uncertain type (agu) that have the integrated photon flux
$F(100 \mbox{ MeV} \le E \le 100 \mbox{ GeV}) \ge 2\times10^{-8}~{\rm ph\,~cm^{-2}~s^{-1}}$. Then, we
excluded sources with $|b|\le10^{\circ}$ in order to avoid any dependence of the sensitivity on the photon
index and minimize polarization induced by the Galactic dust. Finally, applying visibility
constraints and introducing the optical magnitude cut $R\le17.5$, we obtained the unbiased \g--loud
monitoring sample of 62 sources. The~control sample of \g--quiet sources was selected from
Candidate Gamma-Ray Blazar Survey (CGRaBS) sources \citep{Healey2008} that had not been detected by Fermi
at that time (not~included~in~2FGL). We introduced constraints on the 15 GHz mean flux density $\ge
0.06$ Jy and intrinsic modulation index \textcolor{black}{as defined by Richards et al.
\citep{Richards2011}} $m \ge 0.05$. Then, we applied similar visibility cuts and the $R\le17.5$
condition as for the \g--loud sources. Finally, we randomly selected 15~sources that passed all the
cuts. Two out of 15 sources in our \g--quiet sample appeared in the four-year Fermi-LAT (3FGL)
catalog as \g-ray sources. Therefore, we added two more blazars that satisfied the previous criteria,
but were not reported as \g-ray--loud in either 2FGL or 3FGL. Additionally, to the unbiased samples
of \g--loud and --quiet sources, we observed 24 additional "high~interest" sources, which were
not taken into account in population studies where they could bias results \textcolor{black}{due to
their activity being higher than average}. More details on the sample selection are discussed in~Pavlidou~et~al.~(2014)~\citep{Pavlidou2014}.}

\section{Results}

\begin{figure}[H]
 \centering
 \includegraphics[width=0.447\textwidth]{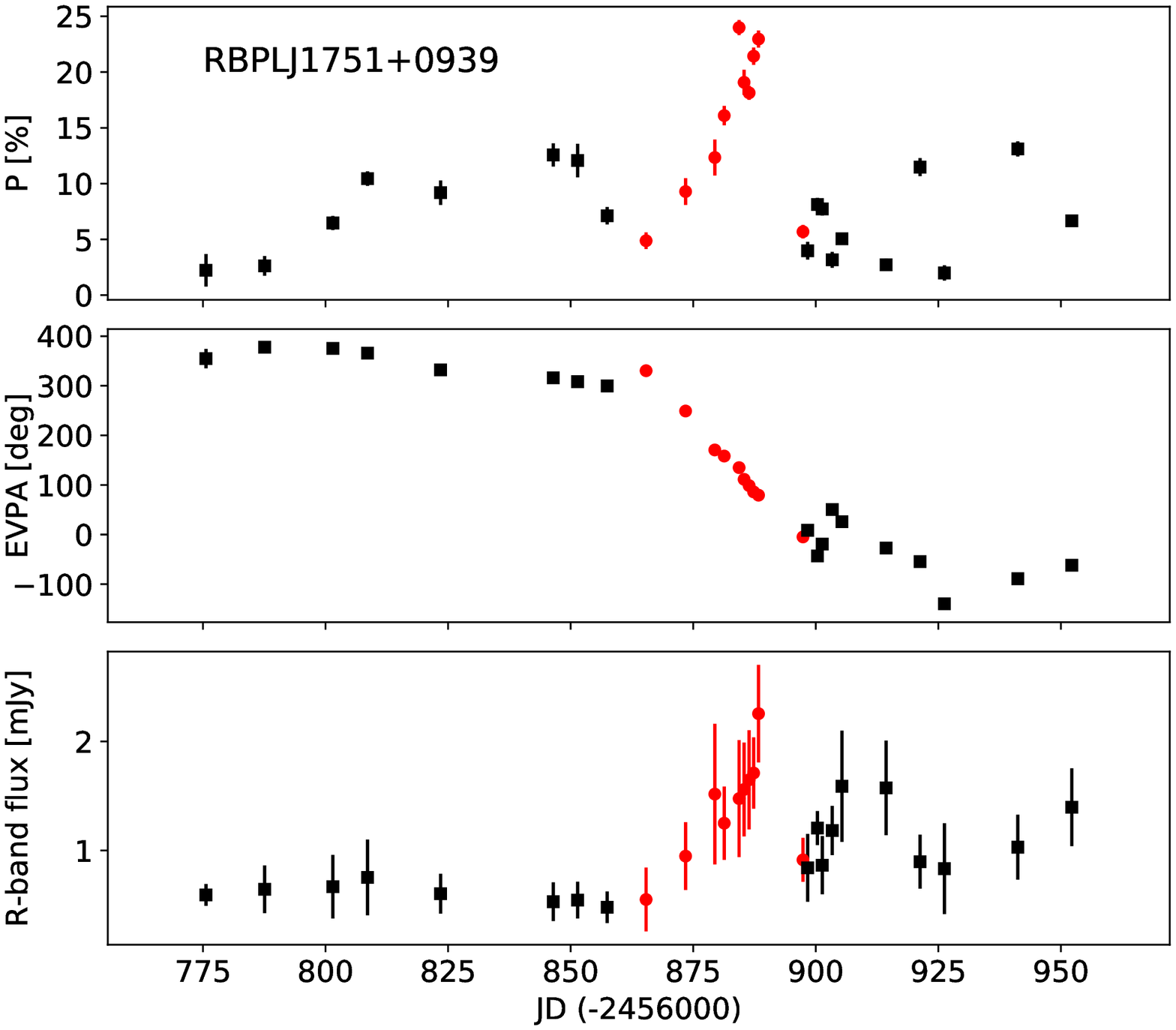}
 \includegraphics[width=0.43\textwidth]{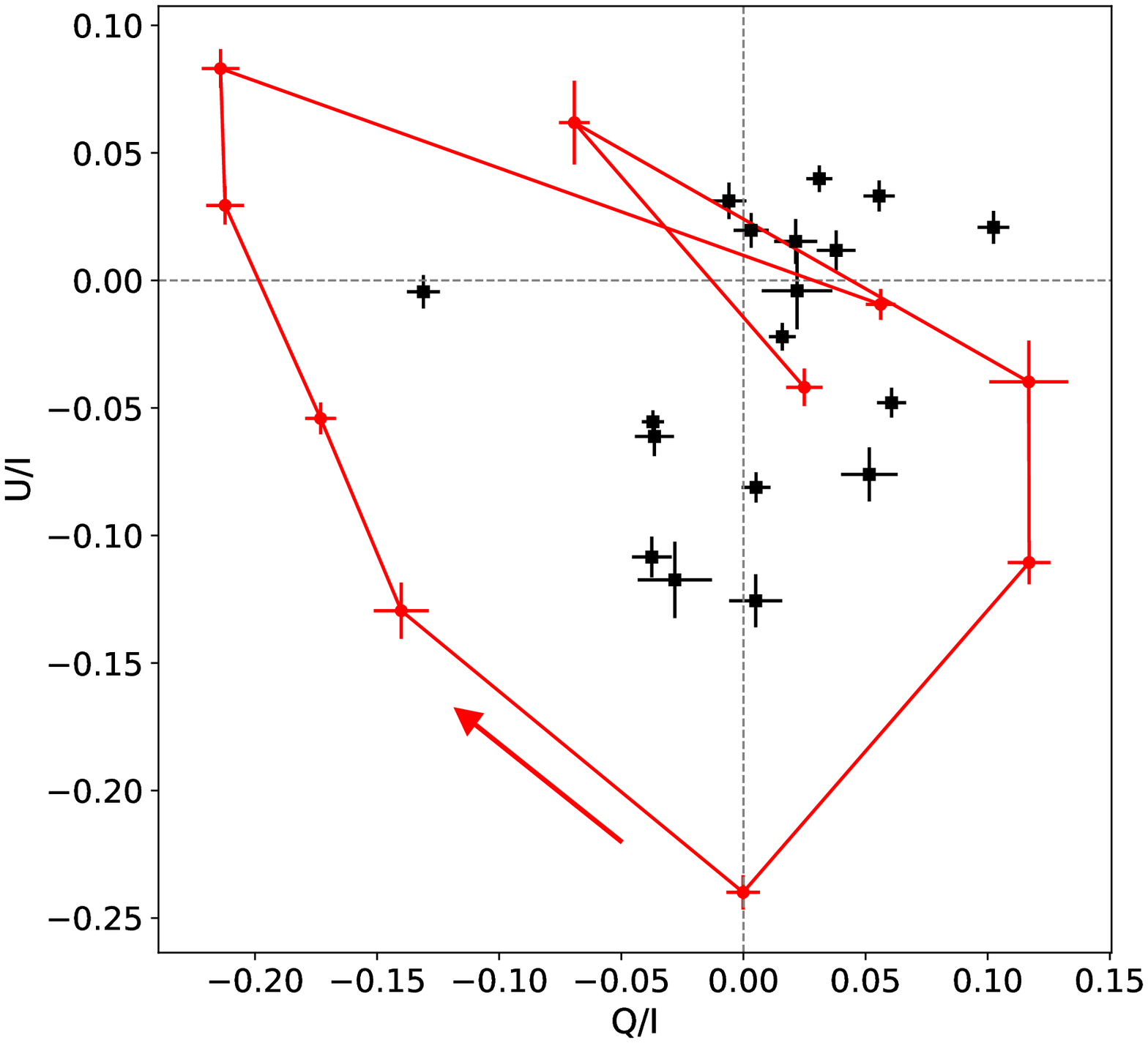}
 \caption{An example of an EVPA rotation (marked by the red circles). Based on data from
\cite{Blinov2016a}.}
 \label{fig:rot_examp}
\end{figure}

After running the program for three observing seasons, we identified 40 EVPA rotations
in 24~individual blazars \textcolor{black}{(see Section~\ref{sec:caveats} for discussion on the EVPA
rotation definition). As an example, in Figure ~\ref{fig:rot_examp} we show one of the rotation events
observed in the \g-ray--loud sample blazar J1751+0939.} We were able to study statistical
properties of the detected events, their characteristics, and connection to the high-energy flares
in blazars. Here, we briefly discuss our major results related to the EVPA rotations, as well as the
behavior of polarization in blazars in general.

\subsection{Polarization Properties of Different Types of Blazars} \label{sec:polbeh}

\textls[-8]{\textcolor{black}{Using data obtained} during the first month of RoboPol operation, we have found that the
average fractional polarizations of the \g--loud and \g--quiet samples are
$<p>$ = $6.4^{+0.9}_{-0.8}\times10^{-2}$ and $3.2^{+2.0}_{-1.1}\times10^{-2}$, respectively. The
hypothesis that the polarization fractions of blazars in these samples are drawn from the same
distribution is rejected at the $3\sigma$ level of significance \cite{Pavlidou2014}. Later, this
result was confirmed using averaged monitoring data for a larger sample of sources
\cite{Angelakis2016}. There, the difference of average intrinsic polarization for the two groups
was found to be even larger, $9.2\pm0.8\times10^{-2}$ and $3.1~\pm0.8~\times~10^{-2}$. A~similar
conclusion that \g-ray luminosity and optical polarization are correlated was made, also based on
Kanata monitoring data \citep{Itoh2016}. These results immediately suggest that the average
``polarizing~efficiency'' in jets of \g--loud and \g--quiet blazars is systematically different.
This efficiency depends on the ordering of the magnetic field in the jet and the electron energy
distribution index. However, the~latter is a rather weak dependence, and cannot solely be responsible
for the wide range of optical polarization observed in blazars (see discussion in
\citep{Itoh2016}).}

As per Hovatta et al. \citep{Hovatta2016}, we tested whether the \textcolor{black}{BL Lac objects} that
were detected with the current-generation TeV instruments had different optical polarization
characteristics compared to similar sources that have not yet been detected. It was found that after
correction for depolarization produced by the host galaxies, TeV-detected and non-detected BL Lacs
had the same mean fractional polarization. The~rate of EVPA variations is also statistically
indistinguishable between the two groups. This was considered as an indication that there are no
intrinsic differences in the polarization properties of the TeV-detected and non-detected
high-energy BL Lac objects. Therefore, the~magnetic field ordering and the electrons distribution
within their jets are not expected to differ systematically.

Besides the relation of optical polarization with \g-ray emission of blazars, it has also been
found that polarization parameters depend on the position of the synchrotron peak maximum. For
instance, low-synchrotron peaked (LSP, $\log(\nu_p) < 14$) sources have higher fractional
polarization and a broader range of values compared to intermediate- (ISP, $14 \le \log(\nu_p)
\le 15$) and high-synchrotron peaked (HSP, $15 < \log(\nu_p)$) sources \cite{Angelakis2016}.
The dependence of the maximum polarization on the synchrotron peak position was also found in
other monitoring programs \citep{Jermak2016,Itoh2016}. Moreover, HSP
 sources tend to demonstrate
\textcolor{black}{significant peaks in EVPA distributions (i.e., a preferred direction of polarization
plane), while LSPs typically have uniform distributions of EVPA.} Hints of this behavior were
noted earlier, but for very small samples of sources so that no
statistically robust conclusion could be made \citep{Villforth2010}.

\begin{figure}[H]
 \centering
 \includegraphics[width=0.5\textwidth]{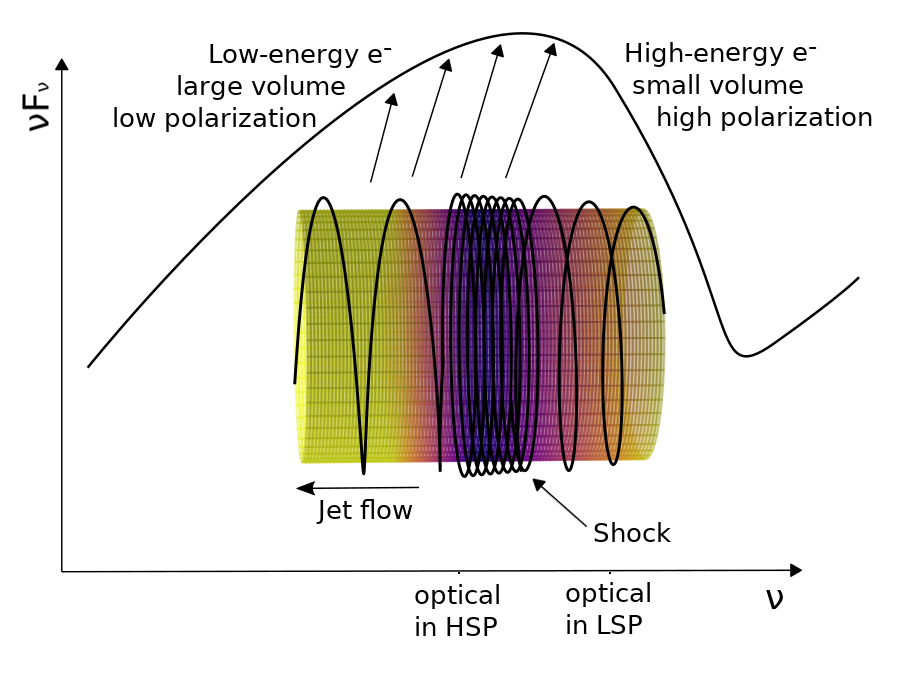}
 \caption{Sketch of the shock-in-jet scenario proposed by Angelakis et al. \cite{Angelakis2016}.}
 \label{fig:jet_sketch}
\end{figure}

Considering the aforementioned dependencies, Angelakis et al. \cite{Angelakis2016} proposed a
qualitative model sketched in Figure ~\ref{fig:jet_sketch}. In this scenario, a mildly relativistic
shock in the jet causes efficient particle acceleration (e.g., due to diffusive shock acceleration
or magnetic reconnection) in a small volume directly downstream of the shock. Propagating further
downstream, these accelerated particles progressively lose their energy due to the emission of
inverse Compton and synchrotron radiation. Therefore, the~emission at the synchrotron peak
frequency or higher is produced by the most energetic particles congregated in a small volume
adjacent to the shock, where the magnetic field is compressed and expected to be highly ordered.
Lower-frequency synchrotron emission is received from a region with greater volume downstream the
jet. This region contains a larger range of magnetic field line orientations. Additionally,
the ordered component of the magnetic field can be less dominant in this zone, when compared to the
turbulent component. Thus, at the fixed frequency of the R-band, we~observe highly polarized
emission near the synchrotron peak maximum and above in LSP sources, and significantly less
polarized emission at the low-energy side of the peak in HSP blazars (see~Figure~\ref{fig:jet_sketch}). Itoh et al. \citep{Itoh2016} also explain the lower polarization of HSP
sources by larger emitting volume and superposition of multiple emitting regions with independent
orientations of magnetic field lines. However, their scenario includes a spine-sheath model, where
the spine has a highly ordered magnetic field and produces highly polarized and variable emission,
while the sheath, in contrast, produces steady and low-polarized light. Then, the systematic difference
in the polarization of HSP and LSP sources can be caused by the increasing jet volume fraction occupied
by the spine. Potentially, an observational test can discriminate between these two models. The
first model predicts that most of the sources \textcolor{black}{must demonstrate decreasing
fractional} polarization from higher (Ultraviolet) to lower frequencies (Near~Infrared), while for
the spine-sheath model, no systematic behavior is expected.

\subsection{Connection of EVPA Rotations and \g-Ray Flares} \label{sec:rot_vs_gamma}

\textls[-5]{One of the main goals of the project was to verify the previously assumed connection of EVPA
rotations with strong \g-ray flares. We addressed this issue in~\cite{Blinov2015} based on the
first-season data and~in~\citep{Blinov2018} using the entire set of 40 events detected during the
three-year program. Our main test included a direct estimation of the probability that all EVPA rotations
occur at random times with respect to \g-ray flares in the light curves of corresponding
blazars.}

\begin{figure}[H]
 \centering
 \includegraphics[width=0.35\textwidth]{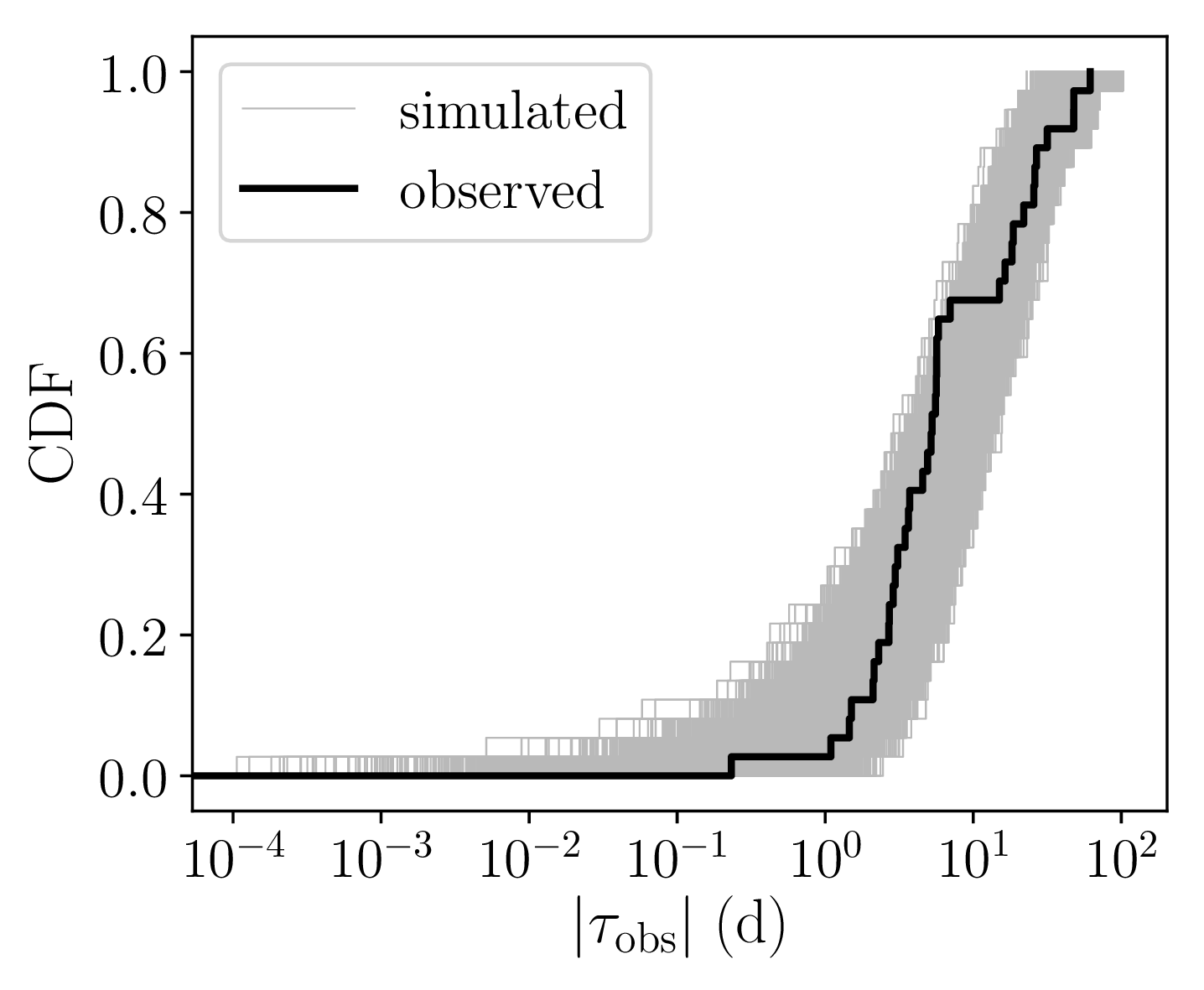}
 \caption{CDFs
 of the time lags $\tau_{obs}$ between the EVPA
 rotations and peak moments of the
closest \g-ray flares for the main sample rotators. The black line represents observed time lags, and the thin grey
lines represent $10^4$ simulated values for the whole sample of rotations. Based on data from
\cite{Blinov2018}.}
 \label{fig:CDF_TL}
\end{figure}

For this purpose, we identified all flares in \g-ray light curves that occurred during intervals
covered by RoboPol. Then, we fitted these flares with an analytic function and derived their basic
parameters, such as the peak time, flux at the peak, rise, and decay times. After this, we found the
time lags between EVPA rotations and flares and constructed the cumulative distribution function
(CDF) of their values (see Figure~\ref{fig:CDF_TL}). Dropping random points at the \g-ray light
curves, we ran a Monte-Carlo simulation that simulated EVPA rotations distributed at random moments
in time. Thus, we~obtained $10^6$ simulated CDFs ($10^4$ of them are shown in
Figure~\ref{fig:CDF_TL}), and found that only 70 were entirely located closer to zero than the observed
one. Therefore, we concluded that the probability of all time lags in the sample being accidentally
so close to zero, as observed, is $\sim 7 \times 10^{-5}$. This implies that at least some of the
time lags between EVPA rotations and \g-ray flare we observed are physically linked.

\begin{figure}[H]
 \centering
 \includegraphics[width=0.36\textwidth]{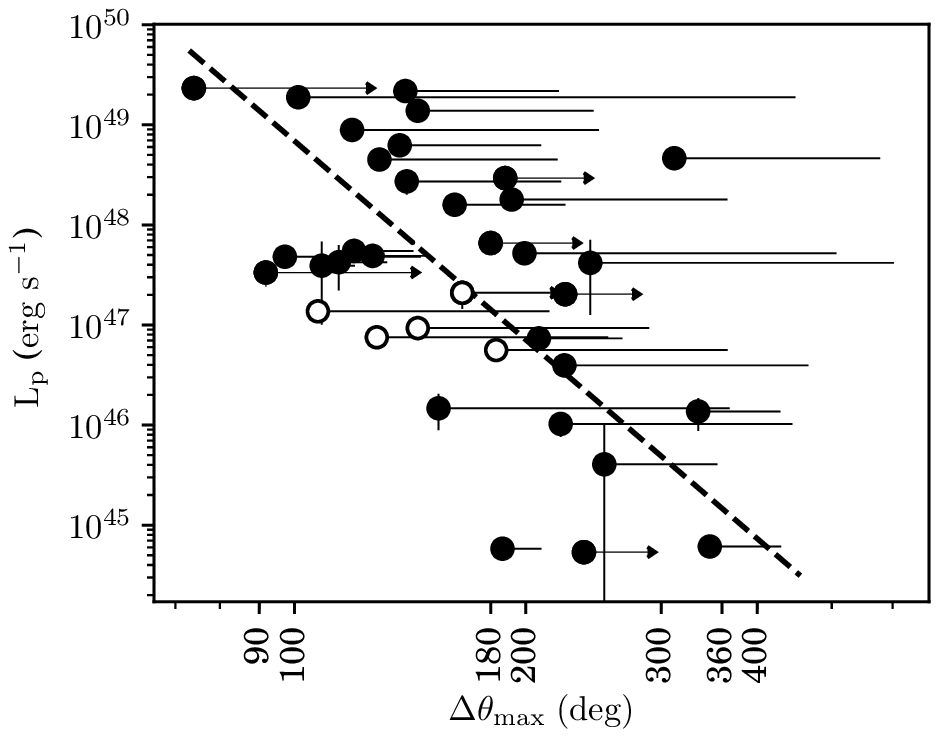}
 \includegraphics[width=0.36\textwidth]{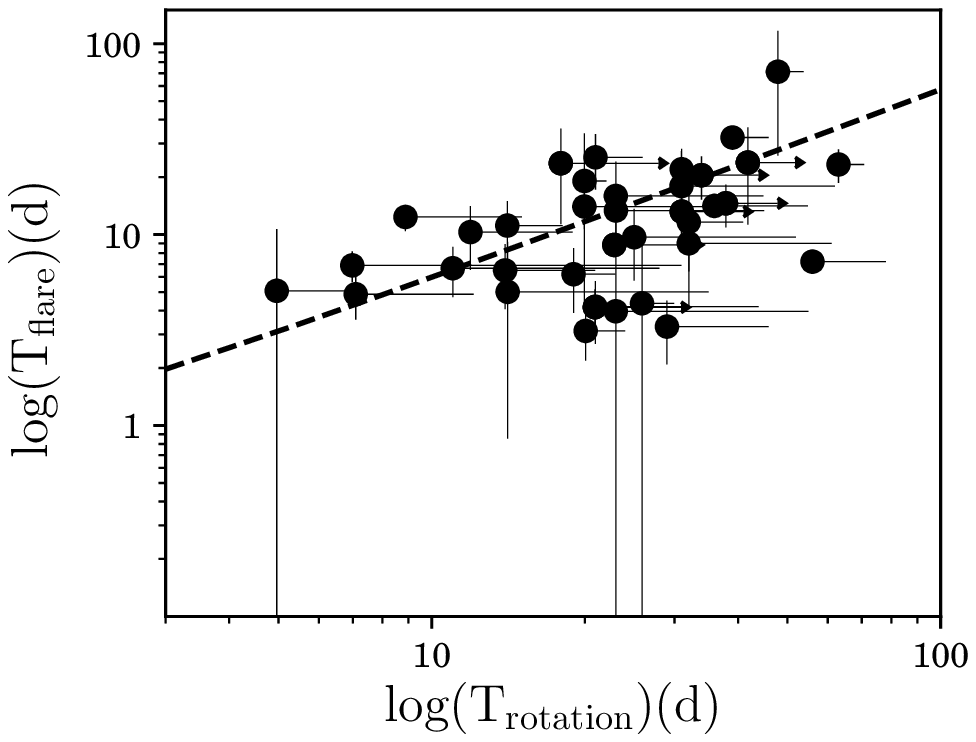}
 \caption{{\bf Left panel}: \g-ray flare luminosity, $L_p$, vs. EVPA rotation amplitude, $\Delta
\theta_{max}$. Empty circles correspond to blazars with uncertain z. {\bf Right panel}: characteristic
timescale of flares, $T_{flare}$, vs. duration of rotations, $T_{rotation}$, in the observer frame
Lines show linear fits to the data. Based on data from \cite{Blinov2018}.}
 \label{fig:gamma_vs_rot}
\end{figure}

\textcolor{black}{This interpretation received further support} when we compared amplitudes and
timescales of EVPA rotations and \g-ray flares (see Figure~\ref{fig:gamma_vs_rot}).
\textcolor{black}{In spite of large uncertainties in determination of the rotations amplitudes
$\Delta \theta_{max}$ (because of the difficulty in identifying the onset and the end of a
particular event), it} has been found that the peak luminosities of flares $L_p$ closest to
rotations are anticorrelated with $\Delta \theta_{max}$. Such a relation could be caused by a
correlation of jet parameters that influence luminosity, i.e., the Lorentz and Doppler factors and
the jet viewing angle, with the rotations amplitudes. The~correlations between these parameters were
indeed found to be significant as well. Moreover, we demonstrated that the characteristic timescales
of rotations and \g-ray flares were marginally correlated \cite{Blinov2018}. Thus, one of the most
significant outcomes of the RoboPol program is that it has proven, for the first time in a
statistically robust fashion, that at least some EVPA rotations must be physically related to \g-ray
flares.

\subsection{Properties of EVPA Rotations}\label{s3.3}

RoboPol provided enough data for a first attempt to characterize the statistical properties of
EVPA rotation events, as well as the behavior of the total flux and fractional polarization during
these events. We detected EVPA rotations with durations, $\Delta T$ from 19 to 518 days in the jet
reference frame and amplitudes, $\Delta \theta$ from 74\dg to 427\dg
\textcolor{black}{(see Figure \ref{fig:gamma_vs_rot})}. The~histogram of rotation amplitudes was found
to be consistent with both normal and uniform distributions. However, it has an apparent peak near
the mean 186\dg value, which is consistent with ``the magic number'' of
180\dg appearing in several models explaining these events. \textcolor{black}{When EVPA rotations are
approximated by straight lines, their} rotation rates $\Delta \theta/\Delta T$ in the jet reference
frame span from 0.2$^{\circ}/{\rm d}$ to 16$^{\circ}/{\rm d}$. They are strongly non-uniform and,
presumably, can be approximated by a power law distribution. This, however, could just be a natural outcome of $\Delta \theta$ and $\Delta T$ being uniformly distributed. The~values
of $\Delta \theta$ and $\Delta T$ are not correlated. We have presented statistical evidence that
$\Delta \theta$ and $\Delta T$ of the rotations are likely to be limited—that is, arbitrary lasting and
high amplitude events are not present in nature. The~null hypotheses that rotation amplitudes and
durations can exceed 460\dg and 500 days are rejected at the significance level of 0.01. These limits
are presumably caused by restraints of the physical parameters presented in the jets, such as
topology of the magnetic field, emission region size, and finite bulk speed of the moving emission
features, or other characteristics related to the EVPA rotations.

We have demonstrated that most of the rotations are accompanied by a decrease in the fractional
polarization. At the same time, its modulation indices, on average, remain constant. Moreover, we~have found a marginal ($2.7\sigma$ significance level) anticorrelation of the depolarization
factor—that is, average polarization during rotation normalized by its value in a non-rotating state,
with the rotation rate in the jet reference frame. This dependence, if confirmed with a larger
sample of events, could place significant constraints on models explaining EVPA rotations.
Finally, it has been demonstrated that during most of the EVPA rotations, there is no clear
systematic increase in the total flux density. These and other properties have been reported in
\cite{Blinov2015,Blinov2016a,Blinov2016b,Blinov2018}.

\subsection{Rotations in Different Types of Blazars} \label{sec:rot_typ}

Thanks to the unbiased nature of the RoboPol monitoring sample, we were able to
investigate the occurrence of EVPA rotations in blazars of different classes \citep{Blinov2016b}. We
started by demonstrating that the polarization plane rotations frequency (at least for events
with rates $<20^{\circ}/d$) is not the same for different sources. For instance, the~null
hypothesis \textcolor{black}{that EVPA rotations occur equally frequently in all sources} was found
to be very unlikely ($p=10^{-7}$). At the same time, this difference in the EVPA rotations
occurrence could not be clearly related to the \g--quiet -- \g--loud dichotomy of blazars. We~could
not reject the null hypothesis that the \g--quiet sample blazars show rotations with the
same frequency as sources in the \g--loud sample. Therefore, according to our data, it is not
impossible that the frequency of EVPA rotations is the same for the blazars in the main and the
control samples. Nevertheless, we found that the frequency of rotations very significantly varies,
even between sources of the main \g--loud sample itself. For example, the~upper limit on the
average frequency of rotation in the \g--loud sample sources, where we detected zero rotations, is
more than two orders of magnitude smaller than frequency of rotations in three sources which exhibited
four rotations during the three years of monitoring. Furthermore, we verified that no external 
reason can entirely explain the distinction in rotation frequencies. A~difference in accuracy of 
EVPA measurements that is caused by various fractional polarization levels cannot be solely 
responsible for this effect. Similarly, we have shown that there is no significant distinction in 
the time-delay factors $\delta/(1+z)$ between the samples with detected rotations and without. 
Therefore, we concluded that there exists a sub-class of objects that exhibit EVPA rotations 
significantly more frequently than others. This difference is not simply related to whether or not a 
blazar is detected by Fermi-LAT. It also cannot be explained by the non-uniformity of observations 
or by observational biases due to differences in the average fractional polarization 
\citep{Blinov2016b}.

\begin{figure}[H]
 \centering
 \includegraphics[width=0.43\textwidth]{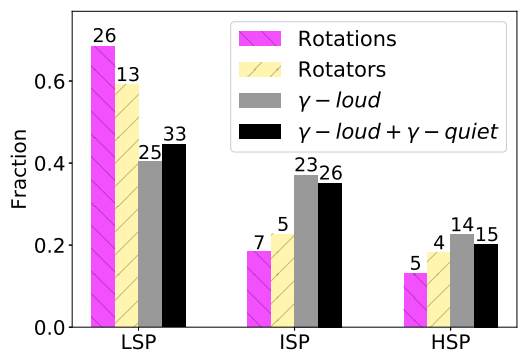}
 \caption{Occurrence of rotations and rotators (sources exhibiting rotations) among blazars of
different synchrotron peak position class. \textcolor{black}{Numbers correspond to the number of
blazars (the number of events in the case of rotations) in each class. Based on data
from \cite{Blinov2016b}.}}
 \label{fig:rot_hist}
\end{figure}

We also looked for other intrinsic reasons that may cause the difference in rotation rates.
{We~compared numbers of EVPA rotations and numbers of sources where these events
occurred (rotators) for the three classes of blazars: low-synchrotron peaked (LSP: $log(\nu_p)<14$),
intermediate- (ISP: $14\le log(\nu_p) < 15$), and
high-synchrotron peaked (HSP: $15 \le log(\nu_p)$).} It was found that rotations tend to occur
in sources with a lower synchrotron peak (see Figure~\ref{fig:rot_hist}). For instance, the~probability
that sources with rotations
were randomly drawn from the parent sample is 1.4\% if both \g--quiet and \g--loud samples
are considered as the parent sample. In the case when we consider only the \g--loud sample, this
probability is reduced to 0.5\%. This tendency of EVPA rotations occurring in LSP blazars is in
agreement with findings of Hovatta et al. \citep{Hovatta2014}, who has shown that LSP blazars are
more variable than HSP in the total optical flux density \textcolor{black}{(Stokes parameter I).}
Their work explains this finding with a scenario similar to the one sketched in
Figure ~\ref{fig:jet_sketch} and discussed before—that is, the LSP sources are observed near and above the
electron energy peak, which causes stronger variations of the emission compared to HSP sources,
where the lower energy electrons cool down slowly and produce mild variability. This scenario is
expected to work for the polarized flux and EVPA variability as discussed in Section~\ref{sec:polbeh}.

We can briefly summarize the major results of the RoboPol project as follows: only about a quarter
of blazars ($\sim$28\% of sources in both \g--loud and \g--quiet samples) exhibit EVPA rotations
with rates $<$$20^{\circ}/{\rm d}$ in the optical band. The~average frequency of appearance of these
events is 1/232 ${\rm d}^{-1}$ (in the observer frame). The~remaining part of sources $\sim$72\% did
not show any rotations registered in our data. Even if they do produce EVPA rotations, this should
happen with average frequency less than $\sim$1/3230 ${\rm d}^{-1}$ in that sample.

\subsection{EVPA Rotations as Random Walks}

The EVPA behaviour in blazars most of the time is consistent with random walk variability
\citep{Moore1982}. This leads to the natural suspicion that long periods of monotonic change of the
EVPA in one direction is nothing but a rare accidental outcome of the same random walk process. We
tested the hypothesis that all EVPA rotations detected during the first observing season of RoboPol
could be explained by stochastic variability of EVPA and found it to be unlikely \citep{Blinov2015}.
In Kiehlmann et al. \citep{Kiehlmann2017} a similarly simple random walk model was tested against
EVPA variability in the three years of monitoring data. They found that the model is able to reproduce
many parameters including durations, amplitudes and rates of EVPA rotations. However, it failed in
simulating realistic distributions of the mean and standard deviation of the fractional
polarization. This potentially could be corrected by more realistic models, such as by introduction of
several populations of sources with different intrinsic variability properties. More comprehensive
models that take into consideration both turbulent and ordered magnetic components as well as
realistic electron energy and density distributions are successful in reproducing the evolution of
polarization parameters and spectral energy distribution in high detail~\citep{Marscher2014}.
RoboPol data do not reject the hypothesis that individual EVPA rotations could be produced by a
random walk of the polarization vector. Nevertheless, the~correlations between parameters of EVPA
rotations and \g-ray flares parameters described in Section~\ref{sec:rot_vs_gamma} would be totally
smeared out if all rotations were produced by random walks. Therefore, we deduced that at least some
rotations are influenced by deterministic processes that define their properties.

It is not impossible that both types (deterministic and stochastic) of EVPA rotation events coexist.
Moreover, they may be present in the emission of a single blazar \citep{Kiehlmann2016}. It could be
that deterministic ones are produced during periods of high activity in the total flux and high
polarization state. While random walks are more likely to detect during low polarization states,
because the uncertainty of EVPA measurements is larger at these periods \citep{Larionov2016}. This
would explain why, on average, we do not see an increase of optical total flux during rotations and
observe depolarization during most of the events.

\subsection{Possible EVPA Rotations in NLSy1 Galaxies}

Radio loud narrow line Seyfert 1 galaxies are an enigmatic type of AGN that host relativistic jets
and have been detected in \g-rays \citep{Foschini2015}. They, presumably, represent the low
luminosity and the low black hole mass tail of the flat spectrum radio-quasars population
\citep{Paliya2018}. Therefore, it is expected that their polarization properties may be similar to
properties of LSP blazars, and they may exhibit high polarization variability together with EVPA
rotations. Indeed, despite a lack of optical polarization measurements, at least some of NLSy1
have demonstrated high fractional polarization variability down to minute scales
\citep{Itoh2013,Eggen2013,Maune2014}.

Angelakis et al. \citep{Angelakis2018} searched for optical polarization plane rotations in emission
of 10 radio--loud NLSy1 galaxies using a combined monitoring data set obtained by RoboPol and other
monitoring programs. They concluded that all sources with \textcolor{black}{multiple measurements
show significant} variability, both in EVPA and fractional polarization. However, due to either
sparse sampling of EVPA curves or high level of noise, they could not confidently report the
detection of a long \textcolor{black}{($>$$90^{\circ}$)} continuous EVPA rotation in NLSy1, despite
having several candidates for such events. \textcolor{black}{Long-term monitoring campaigns} with
more accurate observations are needed to provide clear evidence of similarity of this type of AGN
to blazars in their optical polarization variability.

\section{Caveats with EVPA Ambiguity, Rotation Definition, and Hidden Rotations} \label{sec:caveats}

\begin{figure}[H]
 \centering
 \includegraphics[width=0.64\textwidth]{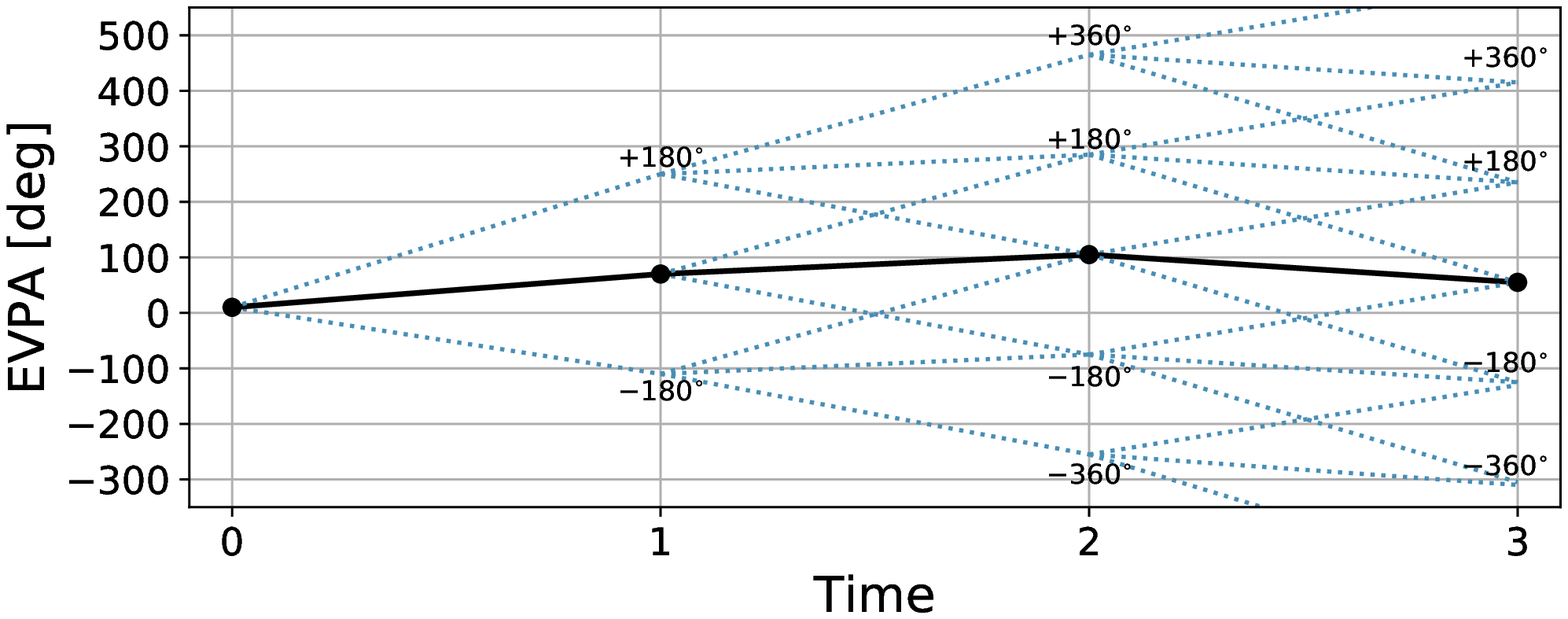}
 \includegraphics[width=0.28\textwidth]{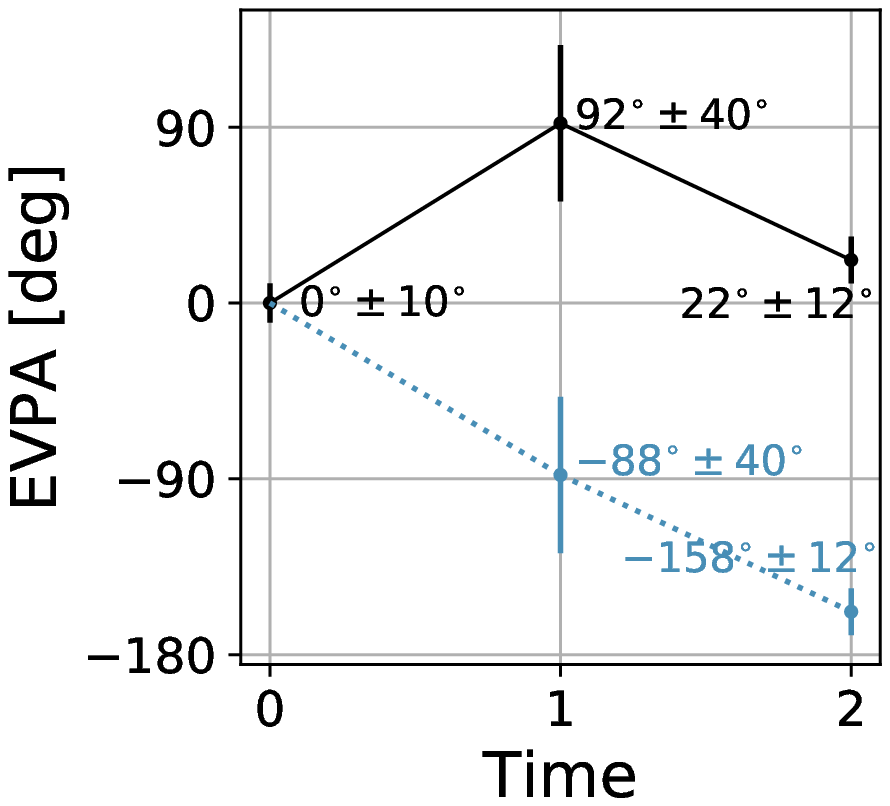}
 \caption{Left panel: an example of a family of EVPA curves \textcolor{black}{demonstrating the
180\dg ambiguity}. All~paths connecting consecutive points in the graph are mathematically
equivalent. Only the three most probable escapes for each step are shown. Right panel: an example of a
spurious jump in the EVPA curve introduced by the 180\dg ambiguity solution method.}
 \label{fig:sim_curve}
\end{figure}

The EVPA defines orientation of the preferred polarization plane with respect to the celestial
meridian. For this reason, the EVPA value is a factor of 180\dg ambiguous, i.e., values EVPA $\pm
n\times180^{\circ}$ are mathematically equivalent. This means that every measurement of EVPA is, in
fact, an infinite countable set of values separated by 180\dg, as is the difference between two
measurements (see the left panel of Figure \ref{fig:sim_curve}). The~most natural way to solve this
180\dg ambiguity is to assume that the variability time scale is much longer than the interval
between measurements \textcolor{black}{(see discussion in Section~\ref{s3.3}~of~\citep{Blinov2015})}
and minimize the difference of EVPA between each of the two consecutive measurements. Nevertheless, 
even the procedure of this minimization can be performed in several different ways. For instance, 
many authors minimize the following value $\Delta \theta_n = |\theta_n - \theta_{n-1}| - SL
\sqrt{\sigma(\theta_n)^2 + \sigma(\theta_{n-1})^2}$, where $\theta_n$ and $\sigma(\theta_n)$ are
$n$-th EVPA measurement and its uncertainty, $SL$ is the significance level typically set to unity.
If $\Delta \theta_n > 90^{\circ}$, the~angle $\theta_n$ is shifted by $\pm n \times 180^{\circ}$,
where the integer $\pm n$ is chosen in such a way that it minimizes $\Delta \theta_n$. If $\Delta
\theta_n \le 90^{\circ}$, then $\theta_n$ is left unchanged.

However, Kiehlmann et al. \citep{Kiehlmann2016} assert that this method itself may lead to
inconsistent results, since it depends on the choice of the initial interval. They suggest to exclude
the term with uncertainties and use $\Delta \theta_n = |\theta_n - \theta_{n-1}|$ for minimization.
Although, in general, these arguments are correct, we caution that this method of the 180\dg ambiguity
solution can lead to spurious $\sim$$180^{\circ}$ jumps in the EVPA curve caused by measurements with
high uncertainties. An example of such a case is shown in the right panel of
Figure~\ref{fig:sim_curve}, where the black and blue EVPA curves are constructed with and without
EVPA uncertainties taken into account.

Besides the high sensitivity of EVPA curves to the particular procedure of the 180\dg ambiguity solution
and sampling, there is \textcolor{black}{ambiguity} in the EVPA rotation definition. Potentially, any
significant change of EVPA between two measurements can be considered as an EVPA rotation. Usually,
long amplitude, and smooth and monotonic changes of EVPA are of interest, and corresponding requirements
are introduced into the definition of an EVPA rotation. \textcolor{black}{For instance, our
definition of an EVPA rotation stated that it must be comprised of at least four measurements and
have a total amplitude $>$$90^{\circ}$, while the start and end points of a rotation event are defined by
a change of the EVPA curve slope by a~factor~of~$\ge$5 or a change of its sign. In the cases with
large gaps $\Delta t > 30$~d between consecutive measurements, we considered rotations being
terminated at these gaps. A~particular} choice of the definition may affect parameters of derived
EVPA rotations. A~different approach for the identification of EVPA rotations was proposed by
Larionov et al. \citep{Larionov2016}, who suggest using the Discrete Correlation Function (DCF)
between the Q and U Stokes parameters. This method, however, requires very
well-sampled light curves that may be unavailable for monitoring programs with large samples and
single observing facilities. \textcolor{black}{For instance, the~EVPA rotations detected by RoboPol
are traced by~4~to~22~measurements, while the DCF method requires an order of magnitude with more dense
sampling.}

Studies of Angelakis et al. \citep{Angelakis2016} and Hovatta et al. \citep{Hovatta2016} demonstrate
that HSP and ISP sources tend to have preferred directions of EVPA, which is presumably caused by an
emission of a steady polarized source superimposed on the erratic component
\citep{Hagen2002,Uemura2010,Villforth2010}. This component leads to an apparent shift of the
measurements distribution centroid $Q_c$,$U_c$ from the origin of coordinates on the Q-U Stokes
parameters plane. In such cases, rotations of the polarization plane may manifest themselves on the
Q-U plane as consecutive loops around ($Q_c$,$U_c$) instead of (0,0). These events show a
low-amplitude variability in EVPA curves and may be missed, since many studies identify rotations
directly in the EVPA curves. Therefore, a~significant fraction of EVPA rotations can be ``hidden'' and
require analysis of the variability in the Q-U plane \citep{Ikejiri2011,Larionov2016b}. A~handy tool
for such analysis was proposed by Uemura et al. \citep{Uemura2016}, who also demonstrated its
efficiency using a combined set of Kanata and RoboPol~observations~\citep{Uemura2017}.

\section{Future Prospects}

Despite the recent progress and advances provided by optical polarimetry monitoring programs, there
are a number of questions related to polarization of blazars that remain unclear. We still do not
know what process or processes produce polarization plane rotations in blazar emission and how they
are related to \g-ray flares. The~short-term polarization variability of AGN remains mostly
unexplored. It is known that both fractional polarization and EVPA can significantly vary on the
minutes–hours time scale \citep{Itoh2013,Itoh2013b,Larionov2016}, but the origin of this variability
is obscure. The~solution of these and other problems requires continuation of systematic optical
polarimetric monitoring programs accompanied by a multiwavelength total-flux follow-up. Moreover,
the necessity of such programs is boosted in the context of future X-ray polarimetry missions
\citep{Fabiani2018,Weisskopf2018}. The~near-future X-ray polarimeters will require rather long
integration times. Therefore, ongoing continuous monitoring of optical polarization in blazars will
be extremely useful for picking highly polarized and active targets for observations in the X-rays.
RoboPol results (see Section~\ref{sec:rot_typ}), as well as theoretical studies \citep{Zhang2013}
suggest that at least HBL sources are expected to be highly polarized and variable in the X-rays
band. Therefore, we expect a bright future for polarimetry of AGN in the optical in synergy with the
X-rays \citep{Peirson2018}.

\vspace{12pt}

\authorcontributions{All authors contributed equally to the manuscript.}

\funding{D.B. acknowledges support by the European Research Council (ERC) under the
European Union’s Horizon 2020 research and innovation program under the grant agreement No 771282.}

\acknowledgments{The {\em RoboPol} project is a collaboration between Caltech in the USA, MPIfR in
Germany, Toru\'{n}~Centre~for~Astronomy~in~Poland, the~University of Crete/FORTH in Greece, and
IUCAA in India.}

\conflictsofinterest{The authors declare no conflict of interest.}



\reftitle{References}






\end{document}